\documentclass[5p,twocolumn]{elsarticle}
\usepackage{subfigure}
\usepackage{amssymb}
\usepackage{amsfonts}
\usepackage{amsmath}
\usepackage{amsthm}
\usepackage{epsfig}
\usepackage{graphicx}
\usepackage[usenames,dvipsnames]{color}
\usepackage[latin1]{inputenc}
\usepackage{booktabs}

\usepackage[hidelinks,unicode=true]{hyperref}
\hypersetup{colorlinks=true,
	linkcolor=blue,
	urlcolor=blue,
	citecolor=blue,
	pdfhighlight=/N
}

\usepackage[normalem]{ulem} 

\newcommand{\av}[1]{\langle{#1}\rangle}
\newcommand{\avqs}[1]{\left[{#1}\right]}

\newcommand{\kmax}{k_\text{max}}
\newcommand{\kmin}{k_\text{min}}

\newcommand{\NumberInfected}{N_\text{I}}
\newcommand{\NumberInfectedExtra}[1]{N_\text{I}^\text{(#1)}}
\newcommand{\NumberISEdges}{N_\text{IS}}
\newcommand{\NumberRecovered}{N_\text{R}}
\newcommand{\NumberIEdges}{N_\text{k}}

\newcommand{\InfectedList}{\mathcal{V}^\text{(I)}}
\newcommand{\InfectedListExtra}[1]{\mathcal{V}^\text{(I,#1)}}
\newcommand{\RecoveredList}{\mathcal{V}^\text{(R)}}
\newcommand{\SusceptibleList}{\mathcal{V}^\text{(IS)}}

\begin{document}
\title{Optimized Gillespie algorithms for the simulation of Markovian epidemic processes on large and heterogeneous networks}
\author{Wesley Cota}
\ead{wesley.cota@ufv.br}
\author{Silvio C. Ferreira}
\ead{silviojr@ufv.br}
\address{Departamento de F\'isica, Universidade Federal de Vi\c cosa,
36570-000, Vi\c cosa, Minas Gerais, Brazil}

\begin{abstract}
Numerical simulation of continuous-time Markovian processes is an
essential and widely applied tool in the investigation of epidemic
spreading on complex networks. Due to the high heterogeneity of the
connectivity structure through which epidemics is transmitted,
efficient and accurate implementations of generic epidemic processes
are not trivial and deviations from statistically exact
prescriptions can lead to uncontrolled biases. Based on the Gillespie
algorithm (GA), in which only steps that change the state are
considered, we develop numerical recipes and describe their computer
implementations for statistically exact and computationally efficient
simulations of generic Markovian epidemic processes aiming at highly
heterogeneous and large networks. The central point of the recipes
investigated here is to include \textit{phantom processes}, that do not
change the states but do count for time increments. We compare the
efficiencies for the susceptible-infected-susceptible, contact process
and susceptible-infected-recovered models, that are particular cases of
a generic model considered here. We numerically confirm that the
simulation outcomes of the optimized algorithms are statistically
indistinguishable from the original GA and can be several orders of magnitude more
efficient.
\end{abstract}


\maketitle

\section{Introduction}

Our daily life is ubiquitously ruled by networked
systems~\cite{Newman10,barabasi2016network,sen2013,Costa2011} as the
transportation infrastructure through which we move, social media in
which we get informed, the biochemical networks regulating the cell
processes inside our bodies, the connections of neurons of our brain,
which are just a few examples. In the last two decades, we have advanced
significantly in the understanding of the structure and functioning of
networks~\cite{Newman10,barabasi2016network}, but we still have
immense challenges in the theoretical frameworks for the dynamic
processes, such as epidemic spreading, evolving on the top of these
networked complex systems~\cite{barabasi2016network,barrat2012}, 
which are mostly large and highly
heterogeneous~\cite{barabasi2016network}. Approximated theories for
the epidemic spreading in complex networks have been intensively
investigated in the last two decades~\cite{PSRMP} and computer
simulations became fundamental tools to corroborate or to point out
the limitations of the theories as well as to provide physical
insights in the construction of new ones. So, accurate and efficient
simulation methods have become imperative for the progress of this
field. Particularly, many real and synthetic networks are very large
and heterogeneous~\cite{Newman10,barabasi2016network,Costa2011}
requiring algorithms which are simultaneously accurate and efficient.

Epidemic models on networks~\cite{PSRMP} assume that individuals of a
population are represented by vertices while the infection can be
transmitted through edges connecting them. Simulations of very large
systems have played a key role in the understanding of central issues
of epidemic spreading on highly heterogeneous networks. Most of these
studies involve thresholds separating an absorbing, disease-free state
and an active phase where epidemics can thrive and, thus, this problem
can be suited in the framework of absorbing-state phase
transitions~\cite{PSRMP,Dorogovtsev08,Marrobook}. The location and the
existence of a finite epidemic
threshold~\cite{Castellano10,Ferreira12,Lee2013,boguna2013nature,
Mata15,Shu,Ferreira16} and the exponents describing the dynamics near
the transition~\cite{Castellano:2006,Hong2007,Castellano08,quenched2011,Mata14} 
have recently been matter of intense discussions fomented by numerical simulations
on networks with power-law degree distributions.

Continuous-time Markovian processes can be simulated using the
statistically exact Gillespie algorithm
(GA)~\cite{Gillespie1,Gillespie2}, and epidemic processes are not
different~\cite{fennell,boguna2014}.  To apply GA to epidemics, one
must decompose the dynamics into independent spontaneous processes and
then perform a change of state by time step that, in turn, is not
fixed. However, every time the state of a vertex changes, the list of all
spontaneous processes must be updated, a task that can be
computationally prohibitive for large networks. Epidemic models are
also frequently implemented using synchronous
schema~\cite{Pastor01,Gomez10,Chakrabarti08,Klemm}, in which all
vertices are updated simultaneously. The time is discretized in
intervals $\delta t$ and the transition rates $\nu$ are converted into
probabilities $p=\nu \delta t$. This prescription is exact  only for
infinitesimal $\delta t$, which is computationally inaccessible, and
large discrepancies with statistically exact versions are observed if
$\delta t$ is not sufficiently small~\cite{fennell,Shu2016}.

In this paper, we discuss statistically exact algorithms derived from
GA for efficient simulations of generic Markovian epidemic processes
on large complex networks. This algorithm has been applied to
investigate the susceptible-infected-susceptible (SIS) (see
Sec.~\ref{subsec:sis_alg}) epidemic model~
\cite{Ferreira12,Lee2013,boguna2013nature,Mata15,Ferreira16, mata2013pair, Cota2016,DeArruda2015a, St-Onge2017} but,
as far we know, an explicit comparison with GA and, therefore, their
statistical exactness has not been assessed. The central difference
between GA and the other algorithms presented here is that the latter permit
steps where no change of configuration takes place, called  of \textit{phantom
processes}, which can hugely reduce the computational time. The
algorithms studied here were conceived to be implemented in diverse
programming languages. Codes in Fortran and Python, which can be
translated to other languages, for the fundamental
susceptible-infected-susceptible (SIS) model were made available in
open access repositories~\cite{siscode}.

We briefly review some basic concepts of network theory and the
analysis of epidemics on finite networks as an absorbing-state phase
transition in sections~\ref{sec:nets} and \ref{sec:abs}, respectively.
Despite of its seminality, the GA is not sufficiently known in
statistical physics and network science communities as we think it
should be. Thus, in section~\ref{sec:GA} we also briefly review the GA
and present the central proofs needed to its derivation that will help
to understand the optimized algorithms to simulate generic epidemic
processes developed in section~\ref{sec:generic}. We further improve
these algorithms to specific epidemic models in
sections~\ref{subsec:sis_alg}, \ref{subsec:cp_alg}, and
\ref{subsec:sir_alg} for SIS, contact process~\cite{Castellano:2006},
and SIR (susceptible-infected-recovered)~\cite{PSRMP}, respectively,
the first two presenting steady active states while the last one does
not. Algorithms for more sophisticated epidemic
processes~\cite{Ferreira16} are presented in
subsection~\ref{subsec:complicated}.  We finally draw our concluding remarks
in section~\ref{sec:conclu}.

\section{Networks as substrates for epidemics spreading}
\label{sec:nets}

In this section, we review some basic concepts of network theory that
are used in the implementation of the algorithms for the simulations
of epidemics. Comprehensive texts can be found
elsewhere~\cite{Newman10,barabasi2016network,barrat2012,Dorogovtsev08}. 
Unweighted and undirected networks are composed by a set of $N$ vertices, 
labeled by $i=1,2,\cdots,N$, and a set of $E$ unordered pairs $(i,j)$ forming the 
edges. Weights and directions can be included through infection rates; see 
section~\ref{sec:generic}. The adjacency matrix is defined as $A_{ij}=1$ if 
there exists an edge connecting $i$ and $j$ (they are neighbors) and $A_{ij}=0$ 
otherwise.

The degree of a vertex $i$ is the number of edges connecting it to
other vertices and is given by $k_i=\sum_j A_{ij}$. The degree of the
most and the less connected vertices of the network are represented by
$\kmax$ and $\kmin$, respectively. The degree distribution 
$P_s(k)$ of a single network realization is
the probability that a randomly selected vertex of the network has
degree $k$ and its moments
\begin{equation}
\av{k^n}=\sum_{k=\kmin}^{\kmax}k^n P_s(k)=\frac{1}{N}\sum_{i=1}^{N}k_i^n
\end{equation} 
are basic quantities that provide valuable statistical properties of
the network. A hub is a vertex with a very large degree compared with
the average: $k\gg \av{k}$. Outliers are vertices, usually hubs, with
degree given by $NP(k)\ll 1$ in an ensemble of networks with  the
expected degree distribution $P(k)$, implying
that only a few of them are observed in a network realization.

Adjacency matrix of networks with a finite average
degree $\av{k}$ are highly sparse and can be efficiently stored and accessed
using the adjacency list~\cite{Newman10}, which is
a one-dimensional array  with $\sum_{i=1}^Nk_i$ elements.
The neighbors of the vertex $i$ are stored between indexes ${p_i}$ and
${p_i+k_i-1}$, where $p_1=1$ and $p_i=1+\sum_{j=1}^{i-1}k_i$ for $i>1$.

There exist several fundamental models of networks
\cite{Newman10,barabasi2016network}. Here, we will consider the
configuration model with a predefined sequence of degrees in which
edges are formed to preserve the degree sequence~\cite{Molloy95}. We
can then elect a form of the ensemble degree distribution $P(k)$,
$k=k_0,\cdots,k_c$ where $k_0$ and $k_c$ are lower and upper cutoffs,
respectively, and the degree of each vertex is a random number
generated according to this distribution~\cite{NR}, forming a
sequence $(k_1,\cdots,k_i,\cdots,k_N)$  of disconnected stubs
in each vertex $i$. Edges are formed by randomly
choosing two stubs and connecting them.

Consider power-law (PL) distributions $P(k)\sim k^{-\gamma}$ bounded
by $k\in[k_0,N]$ where the network size $N$ is large. Using extreme
value theory, one can show that the mean value of largest degree
$\kmax$ scales with network size as~\cite{mariancutofss}
\begin{equation} \av{\kmax} \sim N^\frac{1}{\gamma-1},
\label{eq:avkmax} \end{equation} introducing an average natural cutoff
in the degree distribution. The fluctuations
of $\kmax$ are very large~\cite{Boguna2009} and
$\av{\kmax}$  is not necessarily representative of the maximal
degree of a network realization.

If multiple and self-connections are rejected, the random selection of stubs to connect generates degree correlations in PL
networks with degree exponent $\gamma<3$~\cite{mariancutofss}. A
sufficient condition to eliminate degree correlations in random
networks is to impose a cutoff~\cite{mariancutofss} $k\le
(\av{k}N)^{1/2}$. This result led to the uncorrelated configuration
model (UCM)~\cite{Catanzaro05}, in which the structural upper cutoff
$k_c=\sqrt{N}$ was adopted. For $\gamma<3$, an extreme value theory shows that the fluctuations of
$\kmax$ in UCM networks are relatively small and its value becomes sharply peaked around $k_c$.
For $\gamma>3$, we have that the fluctuations of $\kmax$ diverge as
$N^{1/(\gamma-1)}$~\cite{Mata15,mariancutofss,Boguna2009} implying in large
fluctuations of $\kmax$ and the presence of outliers.

Outliers play important roles on the efficiencies of the simulations.
For SIS, for example, a single outlier can induce a metastable
localized phase~\cite{Mata15,Cota2016} that makes the simulations
computationally much slower. We will discuss the role of outliers in
SIS simulations in subsection~\ref{subsec:sis_alg}. Therefore, for comparisons of
efficiencies  as functions of the network size we will consider a rigid
cutoff defined as~\cite{Dorogovtsev08} $NP(k_c)=1$ such that
$\kmax\sim N^{1/\gamma}$ and presents negligible fluctuations in $\kmax$
also for $\gamma>3$. This cutoff is suitable to evaluate computational efficiencies because the simulation times for different
network samples become essentially the same. Note that this rigid cutoff also renders
uncorrelated networks since $N^{1/\gamma}\ll N^{1/2}$ for $\gamma>2$
and the networks with this cutoff will also be referred as UCM hereafter.

So far we have assumed that the graphs do not evolve,
considering only quenched networks. On the opposite extreme
lay the annealed networks where the connections are
rewired at a rate much larger than the rates of the processes taking place on
them~\cite{Dorogovtsev08}. The adjacency matrix for an annealed
network is defined as the probability that  a vertex $i$ is connected
to  the vertex $j$ in a given time~\cite{Dorogovtsev08}. We associate $k_i$ stubs to the vertex $i$. 
In an uncorrelated model, a stub
can be momentarily connected with any stub of the network with equal
chance including those belonging to the same vertex~\cite{Boguna2009}.
In this model the adjacency matrix becomes
\begin{equation}
\label{eq:adjann}
A^\text{ann}_{ij}=\frac{k_i k_j}{\sum_{l=1}^{N} k_l} = \frac{k_i k_j}{N\av{k}}.
\end{equation}
The computer implementation of uncorrelated annealed networks is simple. We need a
list with $\sum_{i=1}^N k_i$ elements containing $k_i$ copies of each
vertex $i$. The selection of a neighbor of any vertex consists in
choosing at random one element of this list. We will use annealed
networks to introduce epidemics in the realm of absorbing-state phase
transitions in section~\ref{sec:abs}. Algorithms for epidemic
processes on annealed networks are adapted versions of those
for quenched networks and will not be discussed further.

\section{Epidemics in finite networks as an absorbing-state phase transition}
\label{sec:abs}

In closed systems, states where the disease is eradicated are called
absorbing since once one of them is visited, the dynamics remains
frozen in such a state forever. Therefore, a state without infected
individuals is an absorbing one, since infection can only be
produced through interactions involving infected and susceptible
pairs.

\begin{figure}[th]
\centering
\includegraphics[width=0.95\linewidth]{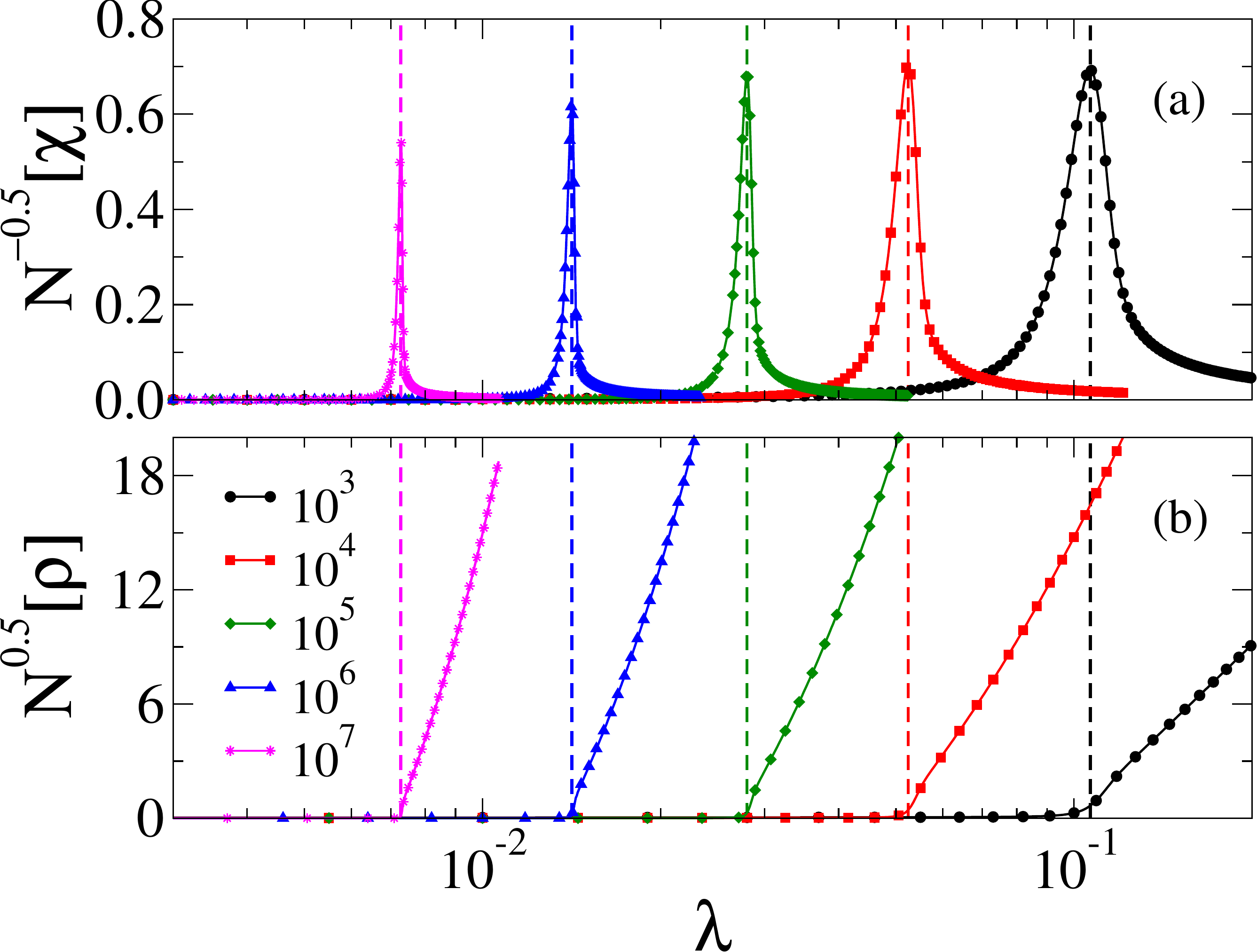}
\caption{Quasistationary (a) susceptibility and (b) density  divided
and multiplied, respectively, by $\sqrt{N}$ (to improve visibility)
against infection rate $\lambda$ for $\mu=1$ and different network
sizes indicated in the legends. The network has a PL degree
distribution with $\gamma=2.3$ and  rigid cutoff $k_c\sim N^{1/\gamma}$.
Note that the susceptibility peak coincides with the point where
density becomes appreciable indicated by dashed lines.
}
\label{fig:chivslb}
\end{figure}

A paradigmatic epidemic model exhibiting an absorbing-state phase
transition is the SIS~\cite{romuvespibook}. The model rules are the
following. Vertices can be infected (denoted by I and $\sigma_i=1$) or susceptible
(S and $\sigma_i=0$). The infected individuals spontaneously heal with  rate
$\mu$. A vertex $i$ can transmit the disease to a susceptible neighbor $j$ with rate
$\lambda_{ij} = \lambda  A_{ij}$, which means that an infected vertex
infects each one of its susceptible neighbors with rate $\lambda$
irrespective of how many it has. This infection rule has deep impacts
on the behavior of the SIS model. A noticeable one is the absence of a
finite epidemic threshold for a PL degree distribution as the network
size goes to infinity~\cite{Castellano10,boguna2013nature}. However,
thresholds of the SIS model can   still be numerically determined for finite
sizes~\cite{Ferreira12,Ferreira16}; see Fig.~\ref{fig:chivslb}.

In order to illustrate  the epidemic spreading as an absorbing-state
phase transition, let us start with the SIS on an arbitrary graph of
size $N$ and adjacency matrix elements $A_{ij}$. The total infection
rate of the network is  given by
\begin{equation}
L=\lambda \sum_{i,j=1}^{N} A_{ij} \sigma_i(1-\sigma_j),
\label{eq:Lsis} 
\end{equation}
while  the total healing rate is
\begin{equation}
M=\mu \sum_{i=1}^{N}\sigma_i = \mu \NumberInfected,
\label{eq:Msis}
\end{equation}
where $\NumberInfected$ is the number of infected vertices.  Now, let us consider the
simple case of an annealed network with a homogeneous degree
distribution~$P(s)=\delta_{s,k}$, that leads to the annealed adjacency
matrix $A^\text{ann}_{ij}=k/N$, which is introduced in
Eq.~\eqref{eq:Lsis} to obtain a total infection rate
\begin{equation}
L=\lambda k \NumberInfected\left(1-\frac{\NumberInfected}{N}\right).
\label{eq:Lsisann}
\end{equation}
Therefore, both total rate of infection and
healing of this model are functions of the number of infected vertices and does not
depend on the specific state.

Let $P_n(t)$ be the probability that there are $n$ infected
individuals at time $t$ and $W_{nm}$ the transition rate from a
state with $m$ to another with $n$ infected vertices. The time evolution of
$P_n(t)$ is given by the master equation~\cite{vankampen}
\begin{equation}
\frac{d P_n}{d t} = \sum_{m=0}^N [W_{nm}P_m - W_{mn}P_n ],
\label{eq:MEOne}
\end{equation}
with rates
\begin{equation}
W_{n+1,n}=g(n)=\lambda k n \left(1-\frac{n}{N}\right)
\label{eq:Wplus}
\end{equation}
and
\begin{equation}
W_{n-1,n}=r(n)=\mu n.
\label{eq:Wminus}
\end{equation}
The transition rates are zero otherwise.

Plugging Eqs.~\eqref{eq:Wplus} and \eqref{eq:Wminus} in \eqref{eq:MEOne}
leads to
\begin{equation}
\frac{d \av{n}}{d t}=\av{g(n)}-\av{r(n)},
\end{equation}
where $\av{f(n)}=\sum_n  f(n)P_n$. Neglecting fluctuations by assuming
that $\av{g(n)}\approx g(\av{n})$, the density of infected vertices
defined by $\rho=\av{n}/N$ evolves as
\begin{equation}
\frac{d\rho}{d t} = (\lambda k-\mu)\rho-\lambda k \rho^2,
\end{equation}
which is easily solved providing the stationary solution
\begin{equation}
{\rho}^* =\left\lbrace
\begin{matrix}
  \frac{\lambda k-\mu}{\lambda k}\sim (\lambda-\lambda_c)^\beta&, &  \lambda\ge\mu/k\\
            0& , & \lambda<\mu/k
\end{matrix}\right.,
\label{eq:rhocpann}
\end{equation}
defining an epidemic threshold $\lambda_c=\mu/k$ and 
an exponent $\beta=1$.

Despite of the approximated solution neglecting fluctuations predicted
an absorbing-state phase transition, one can easily verify that a
stationary (normalized) solution of the master
equation~\eqref{eq:MEOne} with rates given by Eqs.~\eqref{eq:Wplus} and
\eqref{eq:Wminus} is $P_n=\delta_{n,0}$, irrespective of $\lambda$ and
$\mu$. For a finite system the stationary solution is unique if
$W_{nm}$ forms a irreducible matrix~\cite{vankampen}, which is the
present case. So, for finite sizes, this epidemics always visits the
absorbing state at a sufficiently long time since the unique true
stationary state is the absorbing one. This is a universal 
feature of closed systems with absorbing states~\cite{Marrobook}.

Stochastic simulations of epidemic processes can only be performed
with finite systems and  since we do not know the epidemic threshold
\textit{a priori}, we are not able to determine with certainty if the
absorbing state was reached because the infection rate is below the epidemic
threshold or due to the finite size. To handle these difficulties
involving finite sizes and absorbing states, the standard procedure is
to use the standard quasistationary (QS) analysis~\cite{Marrobook}
where the averages at time $t$ are constrained to the samples that did not
visit an absorbing configuration up to time $t$, and then to consider 
a finite-size analysis to extrapolate the thermodynamic limit. Here,
however, we will adopt a simpler method that just prevents the system
from falling into the absorbing state using a reflecting boundary at $n=0$,
in which the dynamics returns to the state that it was immediately
before visiting the absorbing state. The comparison with the
standard QS method in complex networks was recently
performed~\cite{Sander2016}. It has been shown that this method may
provide  finite size scaling exponents different from the standard QS
depending on the network and epidemic model, but preserves the central
properties: Below the epidemic threshold the density of infected
vertices vanishes as $\rho\sim 1/N$ and above it $\rho$ converges to the
actual active stationary value of the infinite system. For the
proposal of this work of comparing the accuracy and efficiency of
distinct algorithms this simpler method is suitable.

The $s$th order moment of the QS density of infected vertices is defined as
\begin{equation}
\avqs{\rho^s} = \frac{1}{N^s}\sum_{n=1}^{N} n^s\bar{P}_n,
\end{equation}
where $\bar{P}_n$ is the QS probability that system has $n$ infected
vertices computed after {a} relaxation time $t_\text{r}$ during an
averaging time $t_\text{av}$. In our notation, $\avqs{\cdots}$
represents the QS averages and we use $\av{\cdots}$ only for the standard ones. 

We will use the dynamical susceptibility~\cite{Ferreira12} defined as
\begin{equation}
\avqs{\chi}=N\frac{\avqs{\rho^2}-\avqs{\rho}^2}{\avqs{\rho}},
\end{equation}
which provides a pronounced peak at the epidemic
threshold for networks without outliers~\cite{Mata15}, as shown in
Fig.~\ref{fig:chivslb}.

\section{The Gillespie Algorithm}
\label{sec:GA}

In general, epidemic Markovian processes on graphs can be composed by
a set of $Z$ independent spontaneous processes $p=1,2,\cdots,Z$ with
rates $\nu_1,\nu_2,\cdots,\nu_Z$ and the probability that the $p$th
process occurs in the infinitesimal interval $[t,t+dt]$ is $\nu_p dt$. For
example, for the SIS model in a state with $\NumberInfected$ infected
vertices and $\NumberISEdges$ edges pointing from an infected to
a susceptible vertex, we can define $Z=\NumberInfected+\NumberISEdges$
independent processes with rates $\nu_p=\mu$ for
$p=1,\ldots,\NumberInfected$, corresponding to transitions
I$\rightarrow$S, and $\nu_p=\lambda$ for
$p=\NumberInfected+1,\ldots,Z$ for the transitions IS$\rightarrow$II.

Spontaneous process is given by the master
equation $\frac{dP_0}{d t}=-\nu P_0$ and $\frac{dP_1}{d t}=\nu P_0$ with
initial condition $P_j(0)=\delta_{j0}$, in which $P_0(t)$ is the
probability that the process has not happen until time $t$ and
$P_1(t)$ is the probability that it has. The solution is
$P_0(t)=e^{-\nu t}$ and $P_1(t)=1-e^{-\nu t}$.

Consider now $Z$ independent spontaneous processes at time $t$. The
probability that the next process is $p$ and that it will happen 
within the interval $[t+\tau,t+\tau+dt]$  is given by
\begin{equation}
\left(\prod\limits_{q=1}^Z e^{-\nu_q\tau}\right)\nu_p dt=\frac{\nu_p}{R} Re^{-R\tau}  dt,
\label{eq:GA1}
\end{equation}
where the product is the probability that no transition has happened
in the interval $[t,t+\tau]$ and $R=\sum_{q=1}^{Z}\nu_q$ is the total rate of
transitions. The right-hand side of Eq.~\eqref{eq:GA1} is suitably
arranged to permit the following interpretation. The next event will
take place after a time $\tau$ with an exponential distribution
\begin{equation}
	P_R(\tau)=Re^{-R\tau}
	\label{eq:PRtau}
\end{equation}
 and this event will be $p$ with probability
$\nu_p/R$. Given the memoryless Markovian nature of the processes, once a
transition has occurred the list of spontaneous processes must  be updated but the
sequence of the dynamics obeys the same rules.

Based on these ideas the GA algorithm is proposed as follows: 1) Build
a list with all spontaneous processes and their respective rates; 2)
Select the time step size $\tau$ from the exponential distribution
$P_R(\tau)$ as $\tau=-\ln(u)/R$, where $u$ is a pseudo random number
uniformly distributed in the interval\footnote{Computationally, the value $u=0$ must be 
strictly forbidden since it leads to an infinity time step.} $(0,1)$; 3) Choose the 
spontaneous process $p$ to take place with
probability $\nu_p/R$, implement it and update the state of the system;
4) Increment time as $t\rightarrow t+\tau$; 5) Return to step 1.
Details of the computer implementation of the GA for epidemics is
given in subsection~\ref{subsec:GAimple}. 
It is worth to stress that GA is a statistically exact method.

Let us exemplify GA with the important problem of the SIS dynamics  on
a star graph, which is composed by $j=1,\ldots,k$ vertices of degree
1, called leaves, connected to a vertex of degree $k$, the center. The
state is determined  by the number of infected leaves and the state of
the center. Let $P(n,\sigma)$ be the probability that there are
$n=0,1,\ldots,k$ infected leaves and the center is in either {the} states
$\sigma=0$ (susceptible) or $1$ (infected). The transitions  and the
respective rates are
\begin{equation}
\begin{matrix}
 (n,0) & \xrightarrow{\lambda n} & (n,1) \\
 (n,0) & \xrightarrow{\mu n}     & (n-1,0) \\
 (n,1) & \xrightarrow{\lambda (k-n)}& (n+1,1) \\
 (n,1) & \xrightarrow{\mu}          & (n,0) \\
 (n,1) & \xrightarrow{\mu n}        & (n-1,1)
\end{matrix}
\end{equation}

\begin{figure}
\centering
\includegraphics[width=0.95\linewidth]{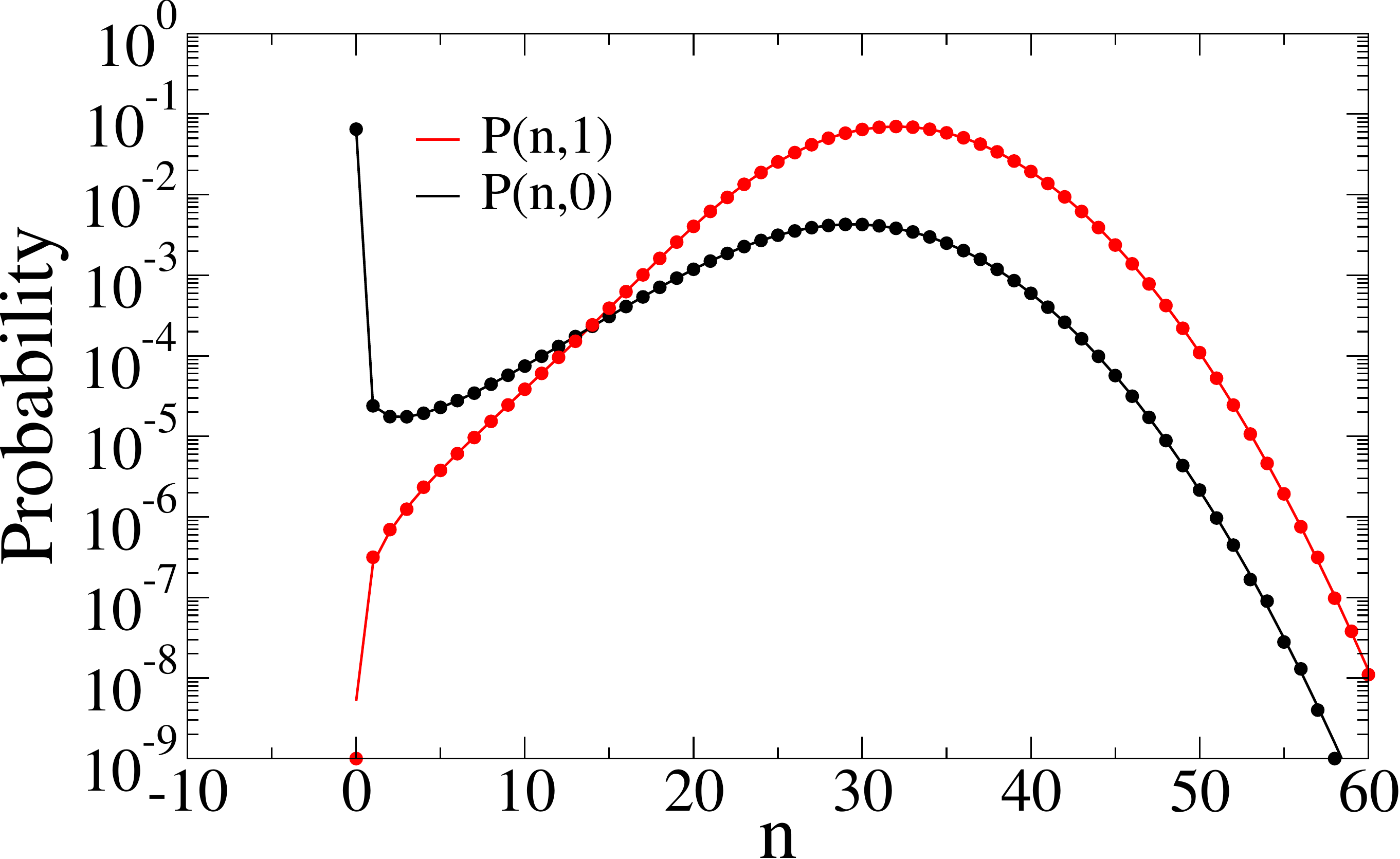}
\caption{Probability distribution for SIS dynamics on a star graph
with $k=100$ leaves at $t=10$ for an infection rate
$\lambda=0.5$ and $\mu=1$. The initial condition is the center infected and all
leaves susceptible. Symbols are GA simulations ($10^7$ samples) and
lines are numerical integrations of the master equation.}
\label{fig:PofnN100lb0p5tmax10}
\end{figure}

The master equation for this process is a set of $2k$ equations that
we numerically integrated using forth order Runge-Kutta  method~\cite{NR}.
This stochastic dynamics can be decomposed into four independent
spontaneous events: a leaf is healed with rate $\nu_1=\mu n$; the center is healed with
rate $\nu_2=\mu\sigma$; the center is infected with rate $\nu_3=\lambda
n(1-\sigma) $; and  a leaf is infected with rate $\nu_4=\lambda
(k-n)\sigma$. In Fig.~\ref{fig:PofnN100lb0p5tmax10}, we compare the
probability distribution $P(n,\sigma)$ obtained using GA simulations
with the numerical integration of the master equation, showing that the
results are completely equivalent, as expected.

\section{Building algorithms for generic epidemic models}
\label{sec:generic}

A generic epidemic dynamics can be modeled by assuming {that} the individuals are in
different epidemiological states~\cite{anderson92}:
susceptible (denoted by S and $\sigma_i=0$)
that can acquire the infection; infected  (I and $\sigma_i=1$) and able to
transmit the disease to a susceptible contact; or recovered
(R and $\sigma_i=2$) representing an immunized state where the individual
cannot neither transmit or acquire the disease.  The generalization to
a higher number of epidemiological compartments is straightforward.
We consider an unweighted network with adjacency matrix
$A_{ij}$. An infected vertex $i$ becomes spontaneously recovered at
rate $\mu_i$, transmits the infection to vertex $j$ with rate
$\lambda_{ij}$ and, if recovered,  turns again to a susceptible  state
(waning immunity~\cite{anderson92}) with rate $\alpha_i$. 

The algorithms require dynamical lists which are represented by
capital calligraphic letters. In both the standard
(subsection~\ref{subsec:GAimple}) and optimized
(subsection~\ref{subsec:opt}) GAs  for the generic epidemic process,
we build and constantly update  two lists $\InfectedList$ and
$\mathcal{M}$ with the positions (labels of the vertices) $\InfectedList_p$ and recovering
rates $\mathcal{M}_p$ of the $p=1,\ldots,\NumberInfected$ infected
vertices. The list updates are simple: the entries of a new infected
vertex are added in their ends. When an infected vertex is chosen
using $\InfectedList$ and  becomes recovered (or susceptible if
$\alpha_i=\infty$), the last entries of the lists are moved to the
index $p$ of the selected vertex, and the list sizes are shortened by
1. Similarly, we build and keep updated the lists $\RecoveredList$ and
$\mathcal{A}$, with the positions $\RecoveredList_p$ and rates
$\mathcal{A}_p$ of the $p=1,\ldots,\NumberRecovered$ recovered vertices. We also
keep updated the total rate  that infected vertices are recovered  and
that recovered ones become susceptible, which are given by
\begin{equation}
M = \sum_{i=1}^{N}\mu_i\delta(\sigma_i,1)=\sum_{p=1}^{\NumberInfected} \mathcal{M}_p 
\label{eq:Mger}
\end{equation}
and
\begin{equation}
A =\sum_{i=1}^{N}\alpha_i\delta(\sigma_i,2)=\sum_{p=1}^{\NumberRecovered} \mathcal{A}_p,
\label{eq:Ager}
\end{equation}
respectively, where $\delta(a,b)$ is the delta Kronecker symbol. The
update of $M$ or $A$ is done by adding (subtracting) $\mathcal{M}_p$
or $\mathcal{A}_p$ when a new element is added (removed) in
$\InfectedList$ or $\RecoveredList$, respectively.

\subsection{Implementation of the Gillespie Algorithm}
\label{subsec:GAimple}

To implement the GA, we need the lists $\SusceptibleList$ and
$\mathcal{L}$, with the positions $\SusceptibleList_p$ of the
susceptible vertices and infection rates $\mathcal{L}_p$ involving the
$p=1,\ldots,\NumberISEdges$ edges connecting infected and susceptible vertices. It
worths to remark that a same susceptible  vertex $i$ will appear $n_i$
times in the list $\SusceptibleList$, where $n_i$ is its number of
infected neighbors. Due to the multiplicity of edges to be
added and deleted from the list every time a change of state occurs, both
$\SusceptibleList$  and  $\mathcal{L}$ are rebuilt and the total infection
\begin{equation}
L  = \sum_{i,j=1}^{N}\lambda_{ij}\delta(\sigma_i,1)\delta(\sigma_j,0)
= \sum_{p=1}^{\NumberISEdges}\mathcal{L}_p
\label{eq:Lger}
\end{equation}
is computed after each event visiting only the
infected vertices and their neighbors with the aid of
$\InfectedList$. 

With these three lists, the steps of GA (\textit{cf}.
section~\ref{sec:GA}) can be implemented as follows. The total rate of
spontaneous processes is $R=M+A+L$. With probabilities $m=M/R$,
$a=A/R$, and $l=L/R$ we choose the class of event I$\rightarrow$R,
R$\rightarrow$S, or IS$\rightarrow$II, respectively. If the event is a
recovering I$\rightarrow$R, one element $p$ of  $\InfectedList$ is
chosen with probability proportional to $\mathcal{M}_p$ and the
respective infected vertex is recovered. If a waning of immunity
R$\rightarrow$S was selected, one element $p$ of $\RecoveredList$ is
chosen with probability proportional to $\mathcal{A}_p$ and the
recovered vertex becomes susceptible. Finally, if an infection event
IS$\rightarrow$II was selected, one element $p$ of $\SusceptibleList$
is selected with probability proportional to $\mathcal{L}_p$ and the
susceptible vertex is infected. The time is incremented by $\tau$
drawn from the distribution $P_R(\tau)$ given by Eq.~\eqref{eq:PRtau}.

The choice of the events proportionally to $\mathcal{M}_p$,
$\mathcal{A}_p$, and $\mathcal{L}_p$  can be implemented using the
\textit{rejection method}, in which an event $p$ is selected with equal
chance and accepted with probability $\nu_p/\nu_\text{max}$ where
$\nu_\text{max}$ is the largest rate in the corresponding kind
of event. The rejection is iteratively repeated until a choice is
accepted. It is simpler and usually more efficient to adopt
$\mu_\text{max}=\max\limits_{i}\lbrace \mu_i \rbrace$,
$\alpha_\text{max}= \max\limits_{i}\lbrace \alpha_i\rbrace$, and
$\lambda_\text{max}=\max\limits_{ij}\lbrace\lambda_{ij}\rbrace$ along
the whole network instead of to update this value after every step.

The implementation of the infection events in this GA follows an edge-based while
the healing and waning of immunity use a vertex-based update scheme. Alternatively,
all events could be performed with vertex-based schemes~\cite{fennell} using a rate
$\nu_i=\mu_i\delta(\sigma_i,1)+\alpha_i \delta(\sigma_i,2) +\delta(\sigma_i,0)\sum_j\lambda_{ji}$. 
This approach, which is not considered in this work,
is computationally slightly simpler to use since no list is necessary. However, its simplest version
demands to visit the whole network to compute the rates after every time step.

\subsection{Optimized Gillespie algorithm (OGA)}
\label{subsec:opt}

\begin{figure}[th]
\centering
\includegraphics[width=0.85\linewidth]{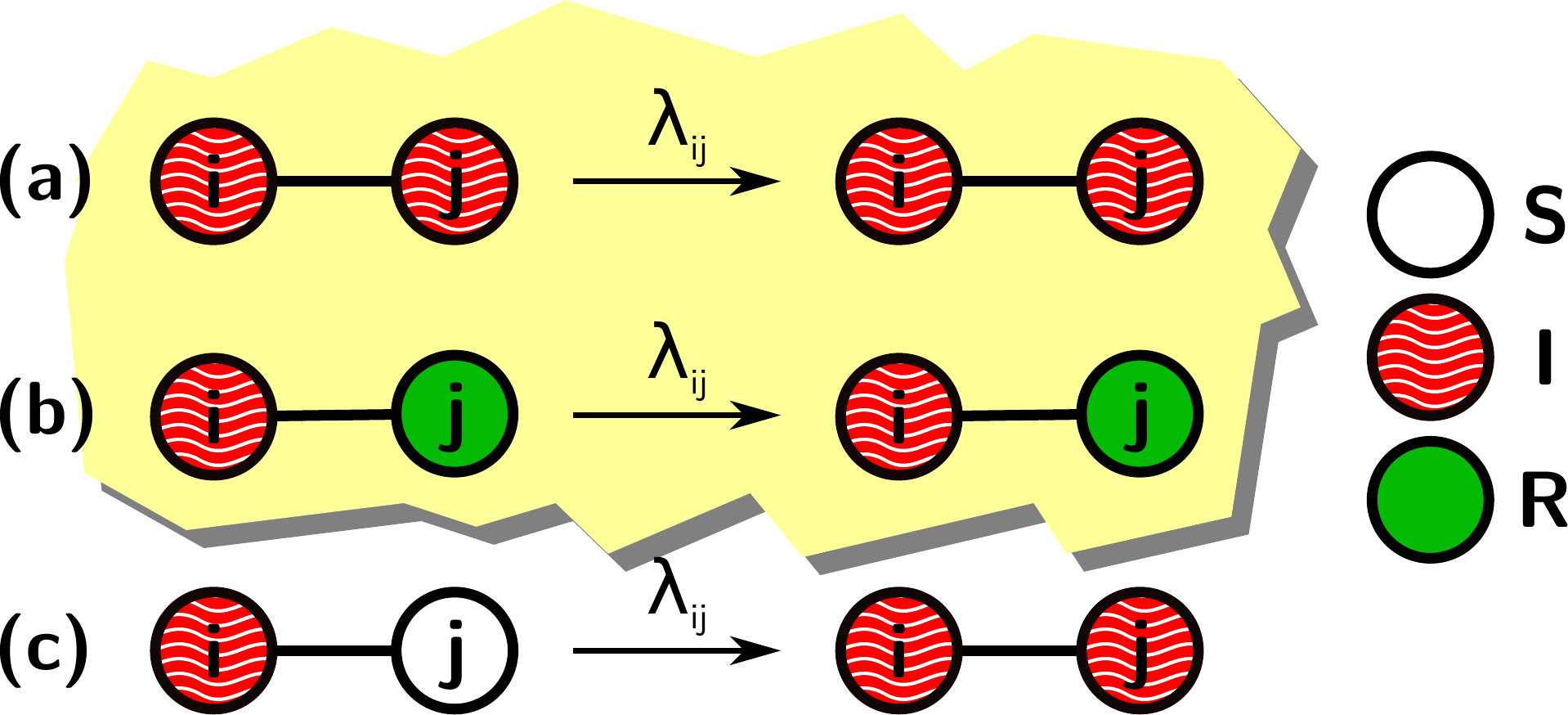}
\caption{Examples of (a)-(b) phantom and (c) real infection processes
for a pair o vertices $(i,j)$.}
\label{fig:phantom}
\end{figure}

Most of the computational load in the original GA holds in building the list
$\SusceptibleList$. Here, we describe a strategy that optimizes this
step by introducing \textit{phantom processes}  that do not change the
state of the system but do contribute for time counting. The phantom
processes here consist of infected vertices $i$ trying to infect other
infected or recovered vertex $j$ with the same rate $\lambda_{ij}$
that they would  infect $j$ if they were susceptible, resulting therefore
in no change of state; see Fig.~\ref{fig:phantom}. We refer to this
algorithm as the optimized Gillespie algorithm (OGA). The method is exact by construction
because it includes processes that are implemented according to the GA rules
but do not change states neither interfere in the processes that actually do.

We write the total infection rate given by Eq.~\eqref{eq:Lger} as
$L=W-P$ where
\begin{equation}
W=\sum_{i,j=1}^{N}\lambda_{ij}\delta(\sigma_i,1)
\label{eq:Qger}
\end{equation}
is the total infection rate emanating from infected vertices, including the phantom processes, and 
\begin{equation}
P=\sum_{i,j=1}^{N}\lambda_{ij}\delta(\sigma_i,1)[\delta(\sigma_j,1)  +\delta(\sigma_j,2)]
\label{eq:Pger}
\end{equation}
is the total infection attempts due to the phantom processes. The total
rate of processes is $R=M+A+W$, which is larger than in GA since $W\ge
L$. 

We need now to calculate once at the beginning of and to keep
stored along the simulation the maximal infection that can be produced by
each vertex, given by
\begin{equation}
w_i=\sum_{j=1}^{N}\lambda_{ij}.
\label{eq:wgen}
\end{equation}

The general epidemic dynamics can be simulated as follows. We build
and keep updated the lists $\InfectedList$, $\mathcal{M}$,
$\RecoveredList$, and $\mathcal{A}$ defined previously. We also need
a list $\mathcal{W}$ with
the infection $\mathcal{W}_p$ produced by the
$p=1,\ldots,\NumberInfected$ infected vertices of {the} list
$\InfectedList$, and the total infection rate
\begin{equation}
	W = \sum_{p=1}^{\NumberInfected} \mathcal{W}_p,
	\label{eq:QgerList}
\end{equation}
produced by these vertices. The update of $\mathcal{W}$ and $W$
follows the same steps of $\mathcal{M}$, $M$,
$\mathcal{A}$, and $A$ described previously instead of the heavier
task of building the list $\SusceptibleList$ and
calculating $L$ in the original GA. This point is essential for the algorithm 
efficiency as discussed in Sec.~\ref{sec:specific}.

With probabilities $m=M/R$ and $a=A/R$ we perform, respectively, a
recovering or waning of immunity as described for the original GA;
\textit{cf}. subsection~\ref{subsec:GAimple}. With probability
$w=W/R$ an infected vertex is randomly selected as an
element $p$ of $\InfectedList$ with probability proportional to
$\mathcal{W}_p$ and let $i=\InfectedList_p$. Next, one neighbor $j$ of
$i$ is chosen with probability proportional to $\lambda_{ij}$ using
the rejection probability
$\lambda_{ij}/\lambda_{i}^{(\text{max})}$, where
$\lambda_{i}^{(\text{max})} =
\max\limits_{j}\lbrace\lambda_{ij}\rbrace$. If the selected neighbor
is susceptible, it is infected. If the selected neighbor is infected
or recovered, \textit{i.e.} a phantom process, no change of state is
implemented, the time is incremented as in the original GA using
$R=M+A+W$, and the simulation runs to the next time step. Note
that the frustrated infection attempts reckon exactly the total rate
of phantom processes $P$ given by Eq.~\eqref{eq:Pger}. The
values of $\lambda_{i}^{(\text{max})}$ need to be computed once at the
beginning of the simulation. 
Depending on the model, further simplifications and improvements can
be adopted. 

\subsection{Improved optimized Gillespie algorithm (IOGA)}

We can improve the rejection method using smarter strategies to reduce
the number of rejections with the cost of storing and updating more
information. We call this method of improved optimized Gillespie
algorithm (IOGA). For epidemic models on networks, the rates can be
very heterogeneous as, for example, the total infection rates produced
by vertices in the SIS model. 

Let us consider the IOGA implementation for the infection processes
using the simplest case with two lists for infected vertices. Define
$w_\text{max} = \max\limits_{i}\lbrace w_i \rbrace$, \textit{cf}.
Eq.~\eqref{eq:wgen}, and let $w_*$ be a threshold such that
$w_i\le w_*$ for the majority of vertices. We define two groups
of vertices with $w_i\le w_*$ and $w_i > w_*$. We build
separated lists $\InfectedListExtra{low}$ and
$\InfectedListExtra{high}$ with the $\NumberInfectedExtra{low}$ and
$\NumberInfectedExtra{high}$ positions of the infected vertices, and
also the lists of total infection rates $\mathcal{W}^\text{(low)}$ and
$\mathcal{W}^\text{(high)}$, of vertices with $w_i \leq w_*$ and $w_i > w_*$, respectively. 
Then, we compute  the total
infection produced by each group $W^{(\text{low})}$ and
$W^{(\text{high})}$. Note that
$\NumberInfected=\NumberInfectedExtra{low}+\NumberInfectedExtra{high}$
and $W=W^{(\text{low})}+W^{(\text{high})}$. When an infection is
selected to happen following the same rules as OGA, with probability
$W^{(\text{low})}/W$ one element $p$ of $\InfectedListExtra{low}$ is
chosen using a rejection probability
$\mathcal{W}_p^\text{(low)}/w_{*}$ while, with probability
$W^{(\text{high})}/W$, one element $p$ of $\InfectedListExtra{high}$
is chosen using a rejection probability
$\mathcal{W}_p^\text{(high)}/w_\text{max}$. The generalization for
healing and waning of immunity is straightforward. The  time
increment is the same as of OGA.

\section{Application for specific epidemic  models}
\label{sec:specific}

\subsection{SIS model}
\label{subsec:sis_alg}

The  implementation of the SIS model with states $\sigma_i=0$ or 1, rates 
$\mu_i=\mu$, $\alpha_i=\infty$, and
$\lambda_{ij}=\lambda A_{ij}$ in the generic dynamics can be
simplified considerably.  Both GA and OGA for SIS do not need the list $\mathcal{M}$ since
$\mathcal{M}_p=\mu$. The original GA reads as follows.  The total
healing and infection rates are $M=\mu \NumberInfected$ and $L=\lambda
\NumberISEdges$, respectively, and $R=L+M$. With probability $m=M/R$
one infected vertex is chosen with equal chance
($\mu_i=\mu_\text{max}=\mu$) using $\InfectedList$ and healed
(I$\rightarrow$S). With probability $l=L/R$, one element of the list
$\SusceptibleList$ is selected with
equal chance ($\lambda_{ij}=\lambda_\text{max}=\lambda$) and the
corresponding susceptible vertex is infected. The time, number of
infected vertices, the list $\InfectedList$,  number of IS edges and the list 
$\SusceptibleList$ are updated.

We have that $\mathcal{W}_p=\lambda
\mathcal{K}_p$, where $\mathcal{K}_p$ is the degree of the vertex
stored in the $p$th entry of $\InfectedList$, is dispensable since if
the vertex $i=\InfectedList_p$ is selected, its degree
$\mathcal{K}_p=k_i$ is known. The total rates of healing and infection
attempts, Eqs.~\eqref{eq:Mger} and \eqref{eq:Qger}, become $M=\mu
\NumberInfected$ and $W=\lambda \NumberIEdges$, respectively, where
$\NumberIEdges=\sum_{ij}A_{ij}\sigma_i= \sum_{i}k_i\sigma_i$ is the
number of edges emanating from all infected vertices. Then, the total
rate is $R=\mu \NumberInfected+\lambda \NumberIEdges$. With
probability $m=M/R$ an infected vertex $i$ is selected with equal
chance using the list $\InfectedList$ and healed (I$\rightarrow$S).
With probability $w=W/R$, an infected vertex $i$ is selected using
$\InfectedList$ proportionally to its degree with a rejection
probability $k_i/\kmax$.  The same infection rate for all edges
implies that a neighbor of $i$ is chosen with equal chance and, if
susceptible, it is infected. The time, number of infected vertices,
edges emanating from them and the list $\InfectedList$ are updated.

The probability $k_i/\kmax$ of the OGA for SIS implies in
too many rejections for power-law degree distributions.  The fraction of
vertices with degree less than $k_*$ is approximately given by
\begin{equation}
\label{eq:kast}
\sum\limits_{k=k_0}^{k_*}P(k)\simeq \left[1-\left(\frac{k_0}{k_*}\right)^{\gamma-1}\right].
\end{equation}
The optimal choice of $k_*$ will depend
on the degree distribution. To be effective, $k_*$ should
not be much larger than the degree of the wide majority of vertices and thus the result of Eq.~\eqref{eq:kast} must not be far from 1. For IOGA simulations presented here, $k_*=
2\av{k}$ was used.

The IOGA implementation for SIS is the following.  The infected
vertex to be healed is chosen at random from either the lists
$\InfectedListExtra{low}$ or $\InfectedListExtra{high}$ with
probabilities $\NumberInfectedExtra{low}/\NumberInfected$ or
$\NumberInfectedExtra{high}/\NumberInfected$, respectively, since
$M^\text{(low)} = \mu\NumberInfectedExtra{low}$ and  $M^\text{(high)}
= \mu\NumberInfectedExtra{high}$. In the infection event, with
probabilities $\NumberIEdges^{\text{(low)}}/\NumberIEdges$ or
$\NumberIEdges^{\text{(high)}}/\NumberIEdges$, one vertex $i$ is
selected from $\InfectedListExtra{low}$ or $\InfectedListExtra{high}$
using the rejection probabilities $k_i/k_*$ or $k_i/\kmax$,
respectively.  The time, number of infected vertices in each compartment
($\NumberInfectedExtra{low}$ and $\NumberInfectedExtra{high}$), edges
emanating from them ($\NumberIEdges^{\text{(low)}}$ and
$\NumberIEdges^{\text{(high)}}$) and the lists
($\InfectedListExtra{low}$ and $\InfectedListExtra{low}$) are updated.

\begin{figure}[hbt]
 \centering
 \includegraphics[width=0.95\linewidth]{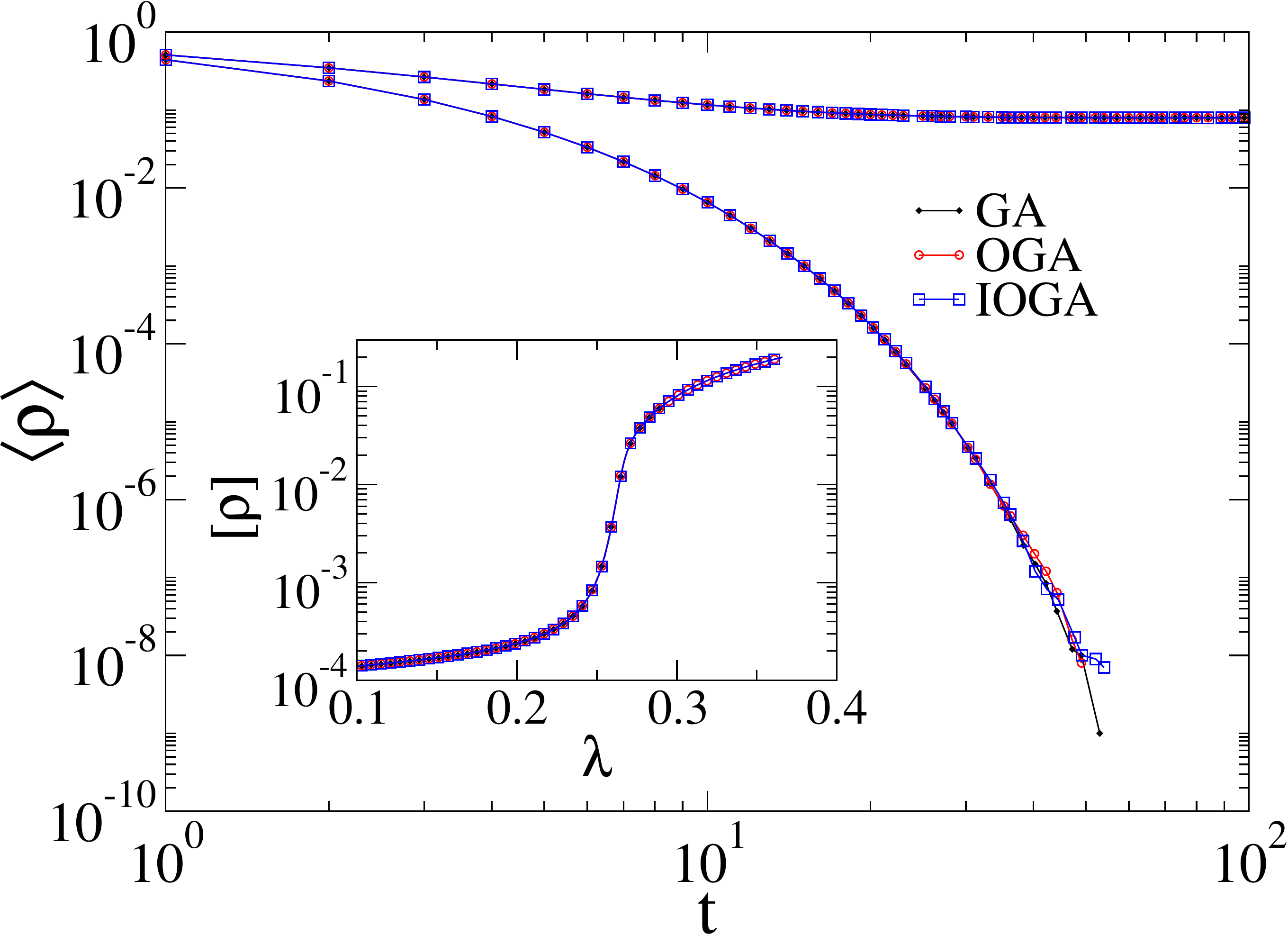}
 \caption{Density against time for SIS model with $\mu=1$ on a single
 network realization of the UCM model with $\gamma=4.0$, $k_0=3$,
 $N=10^4$, and $k_c\sim N^{1/\gamma}$. The infection rates are
 $\lambda=0.150<\lambda_c$ (bottom curves) and
 $\lambda=0.300>\lambda_c$ (top curves). The averages were done  over
 $10^5$ and $10^4$ independent runs below and above the epidemic
 threshold with initial condition $\rho(0)=1$. Inset shows the QS
 density against infection rate.}
 \label{fig:decaySIS}
\end{figure}

\begin{figure}[ht!]
 \centering
 \includegraphics[width=0.90\linewidth]{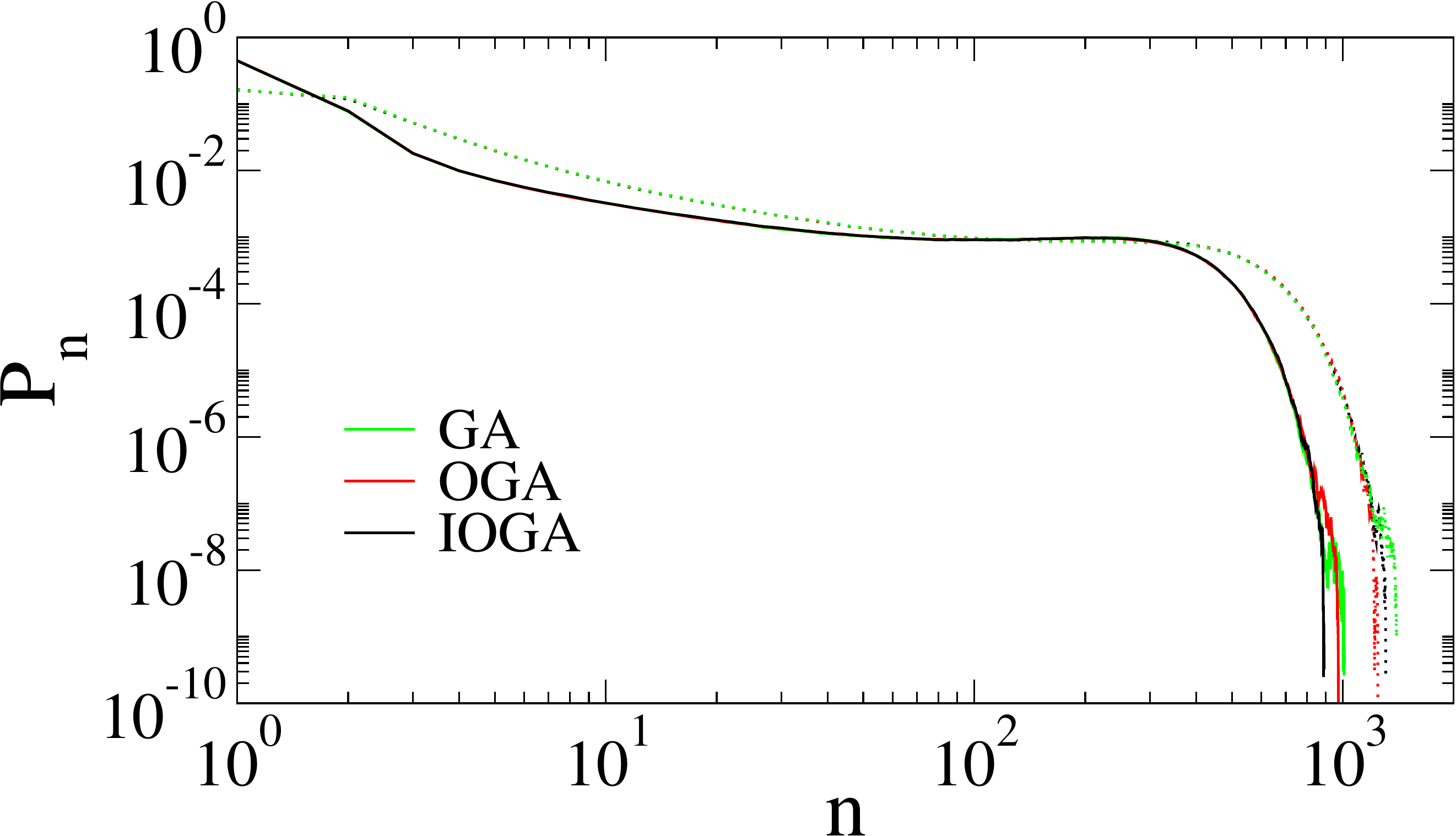}
 \caption{Quasistationary probability distribution for SIS model at
 the epidemic threshold	estimated via susceptibility (see Table~\ref{tab:thre}) with $\mu=1$ on
 UCM networks using distinct algorithms. The network
 parameters are $N=10^5$, $k_0=3$, $k_c\sim N^{1/\gamma}$,
 $\gamma=2.3$ (solid lines)  and $\gamma=4.0$ (dotted lines).
}
 \label{fig:histo}
\end{figure}

Typical outcomes of different algorithms for SIS simulations are
compared in Fig.~\ref{fig:decaySIS} for a single realization of a same
UCM network with exponent $\gamma=4.0$, $N=10^4$ and cutoff $k_c\sim
N^{1/\gamma}$. In the main plot, we show the density of infected
vertices against time using standard averaging (samples that visited
absorbing states before time $t$ are reckoned) over many independent
runs with initial condition $\rho(0)=1$. In the inset, the QS density
against infection rate is shown. Differences between curves are
noticeable only for very low densities due to the finite number of
samples. Excellent matches are also obtained for the QS probability
distribution, a benchmark of the dynamics, of SIS epidemics at the
threshold as shown in Fig.~\ref{fig:histo}. Tiny differences due to
finite statistics are present for very low probabilities. Essentially
perfect matches of the QS distributions are also found in both super
and subcritical phases. Fortran and Python codes for the 
decay simulations are available in~\cite{siscode}.
The former  was tested for GNU~\cite{gfortran} and
Intel~\cite{ifort} non-commercial Linux versions of Fortran 
and the latter using Python 3.6.0~\cite{python}.

Here, it is worth to mention that the average time step given by
$\av{\tau} =1/R$  is commonly used as the time
increment~\cite{Marrobook,Ferreira12,Lee2013,boguna2013nature,Mata15,Castellano:2006,DeArruda2015a,St-Onge2017} 
instead of drawing it according to an exponential distribution given by Eq.~\eqref{eq:PRtau}. 
We verified that this step of the implementation is irrelevant for both QS analysis and decay 
simulations with large averaging and thus these previously obtained results are computationally valid.

We compared the CPU times performing QS simulations at and above the
epidemic threshold estimated via the maxima of the susceptibility for
$\mu=1$; see Table~\ref{tab:thre} for the threshold values used in the simulations. We started with 1\% of infected vertices and ran the dynamics
during {$t_\text{av} + t_\text{rlx} = 3\times 10^6$}. Networks with
different levels of heterogeneity were investigated: weakly heterogeneous
networks with degree given by $k=k_0+k'$, where $k_0=3$ is fixed and
$k'$ is drawn from an exponential distribution $P(k')=a\exp(-k'/a)$
with $a=3$ and $k'\ge 0$; UCM networks with either  $\gamma=2.3$ and
$4.0$,  $k_c\sim N^{1/\gamma}$ and minimal degree $k_0=3$. {All CPU
	time comparisons were performed in a workstation with two six-core
	Intel Xeon processors E5-2620 2.00 GHz and 32 Gb of RAM memory}. The
code was written in {Fortran} and compiled with a non-commercial
version of the Intel Fortran for Linux 64-bit using double precision and
standard compilation optimizations.

\begin{table}[htb] 
	\begin{center}
		\setlength{\tabcolsep}{0.45em}
		{\normalsize\renewcommand{\arraystretch}{1.2}
			\begin{tabular}{@{}llllll@{}}
				\toprule
				\multicolumn{1}{l}{$N$} & \multicolumn{3}{l}{SIS} &
				\multicolumn{2}{l}{CP} \\ 
				\cmidrule(lr){2-4}%
				\cmidrule(lr){5-6}%
				\multicolumn{1}{l}{} &Exp. & $\gamma = 2.3$ &$\gamma = 4.0$ & $\gamma = 2.3$ &$\gamma = 4.0$ \\
				\midrule
					$1\times10^3$ & $0.1705$ & $0.1065$  & $0.3113$ & $1.2606$  & $1.4691$ \\ 
					$3\times10^3$ & $0.1674$ & $0.0749$  & $0.2912$ & $1.2181$  & $1.4338$  \\ 
					$1\times10^4$ & $0.1614$ & $0.0524$  & $0.2619$ & $1.1807$  & $1.4058$  \\ 
					$3\times10^4$ & $0.1605$ & $0.0393$  & $0.2515$ & $1.1626$  & $1.3950$  \\ 
					$1\times10^5$ & $0.1603$ & $0.0280$  & $0.2319$ & $1.1445$  & $1.3872$  \\ 
					$3\times10^5$ & $0.1602$ & $0.0198$  & $0.2254$ & $1.1335$  & $1.3855$  \\ 
					$1\times10^6$ & $0.1593$ & $0.0141$  & $0.2166$ & $1.1255$  & $1.3830$  \\ 
					$3\times10^6$ & $0.1592$ & $0.0104$  & $0.2065$ & $1.1197$  & $1.3813$  \\ 
					$1\times10^7$ & $0.1591$ & $0.0073$  & $0.1956$ & $1.1145$  & $1.3807$  \\\bottomrule
		\end{tabular}}
	\end{center}
	\caption{Threshold values used in the simulations of SIS and CP dynamics with $\mu = 1$ for networks with different levels of heterogeneity, estimated via susceptibility.}
	\label{tab:thre}
\end{table}

\begin{figure}[ht!]
 \centering
 \includegraphics[width=0.99\linewidth]{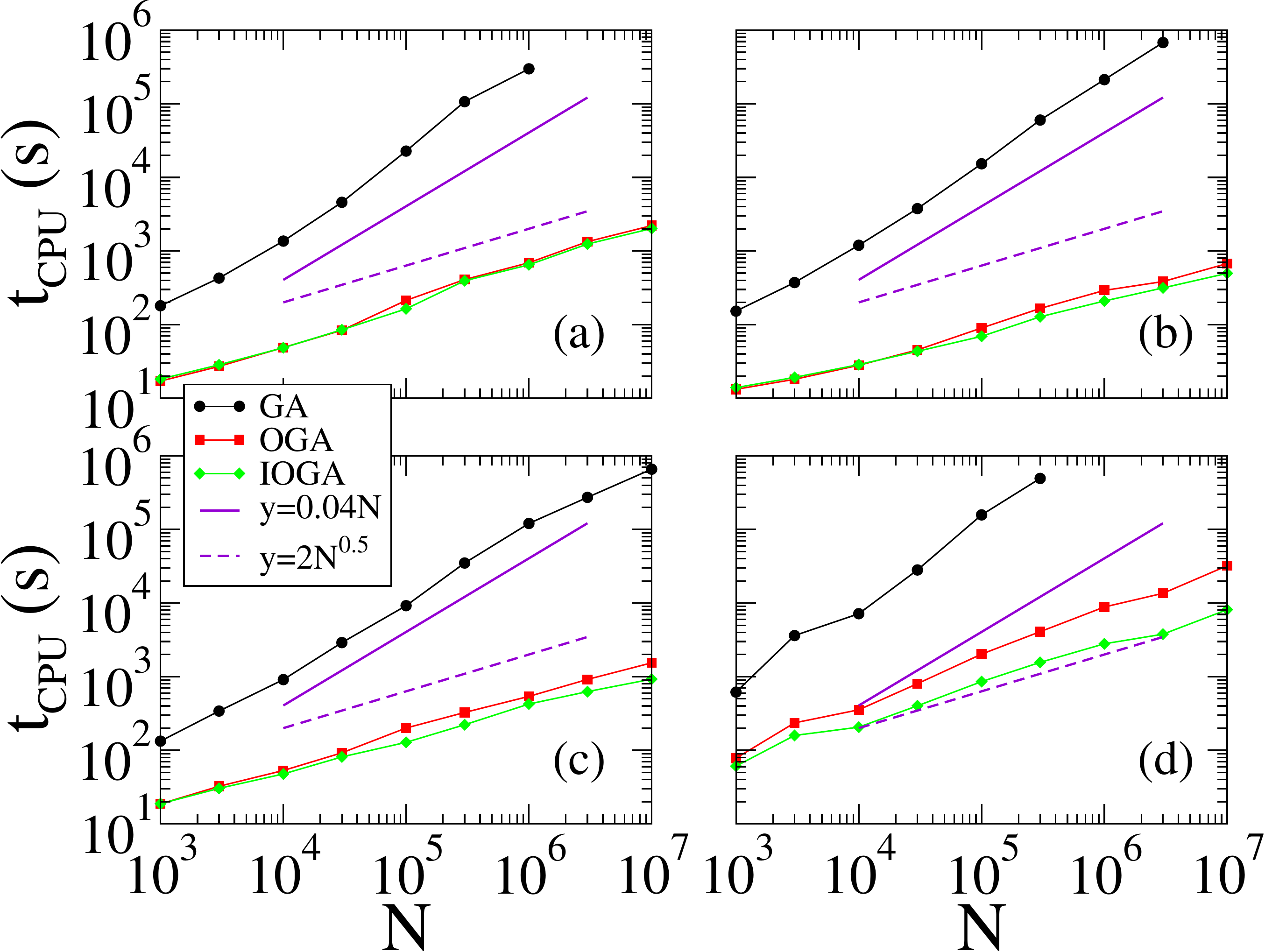}
 \caption{CPU times for QS simulations of the SIS dynamics at the epidemic
 threshold  (see Table~\ref{tab:thre}) for networks with different levels of heterogeneities: (a)
 Exponential; (b) UCM with $\gamma=2.3$; (c) UCM with $\gamma=4.0$;
 and (d) UCM with $\gamma=4.0$ plus an outlier (see main text). The total
 simulation time is $t_\text{av}+t_r=3\times 10^6$. Solid and dashed
 lines are guides to the eyes.}
 \label{fig:cpusiscrit}
\end{figure}

CPU times for the SIS at the epidemic threshold as a function of
the network size  are compared in
Fig.~\ref{fig:cpusiscrit} and Table~\ref{tab:sis}.  The CPU
time for GA increases almost linearly while for OGA and IOGA it does
sublinearly. The relative gain of IOGA in comparison with OGA  is 
appreciable even in networks without outliers as can be seen in
Table~\ref{tab:sis}.

\begin{table}[htb] 
 \begin{center}
 	\setlength{\tabcolsep}{0.4em}
{\normalsize\renewcommand{\arraystretch}{1.2}
 \begin{tabular}{@{}lllllllll@{}}
 	\toprule
\multicolumn{1}{l}{} & \multicolumn{2}{l}{GA} &
\multicolumn{3}{l}{OGA} & \multicolumn{3}{l}{IOGA}  \\ 
\cmidrule(lr){2-3}%
\cmidrule(lr){4-6}%
\cmidrule(lr){7-9}%
\multicolumn{1}{r}{$N$} &$10^5$ &$10^6$& $10^5$  &$10^6$& $10^7$& $10^5$ & $10^6$& $10^7$ \\
\midrule
Exp.             			& 379 	& 4970 	& 3.55 	& 11.6 &	36.9		& 2.73  & 10.8 &	33.7		\\
$\gamma=2.3$    & 255 	& 3540 	& 1.50 	& 4.86 &	11.3		& 1.15  & 3.48 &	8.27		\\
$\gamma=4.0$    & 153 	& 2010 	& 3.34 	& 8.99 &	25.7		& 2.14  & 7.09 &	15.4		\\
Outlier          		 & 2630	& ---  		& 33.8 	& 147  &	536		& 14.3  & 46.6 & 	136	\\\bottomrule
 \end{tabular}}
 \end{center}
\caption{CPU times in minutes for QS simulations of the SIS model at
	the epidemic threshold in networks of different sizes and levels of
	heterogeneity using different simulation methods.  The total
	simulation time is $t_\text{av}+t_r=3\times 10^6$ steps.}
\label{tab:sis}
\end{table}

To investigate the role of outliers in the computer efficiency we
considered the UCM network with $k_c\sim N^{1/\gamma}$ adding a vertex
with degree $k_\text{out}\gg k_c$. We chose
$k_\text{out}=\av{\kmax}$ where $\av{\kmax}$ is the mean value of
the maximal degree obtained in network with cutoff $k_c=N$. The
presence of this outlier leads to a metastable and localized phase in
SIS dynamics: see also Refs.~\cite{Mata15,Cota2016,DeArruda2015a}.  We
performed simulations for $\gamma=4.0$ at the epidemic threshold
determined for the network without the outlier, shown in Table~\ref{tab:thre}. The system having an
outlier remains critical with density of infected vertices decaying
as $\rho_\text{out} \approx 20N^{-0.67}$, which  is much
larger than the density obtained  without the outlier  given by
$\rho\approx 3.4N^{-0.67}$. Figure~\ref{fig:cpusiscrit}(d) shows the
CPU times obtained for simulations with the outlier using different
methods. The computational gain of IOGA in relation to OGA is very expressive and
becomes more relevant as the network size increases as can be also seen 
in Table~\ref{tab:sis}.

Above the epidemic threshold all simulations becomes much  slower than
the critical ones. We compared the efficiencies for a UCM network with
$\gamma=2.3$ and $k_c\sim N^{1/\gamma}$ at an infection rate
$\lambda=\lambda_c+0.02$ and $\mu=1$. The CPU time increases linearly
with size for OGA and IOGA and no significant difference between them
was observed in the absence of outliers. {The GA simulations becomes
	exceedingly slow {with CPU times} scaling as $t_\text{CPU}\sim
	N^{1.6}$}. For example, the QS simulation in networks with $N=10^4$
and  $t_\text{av}+t_r=3\times 10^6$  takes approximately {2.5 days} for GA against 10
min for OGA or IOGA.

The slowness of GA is due to the building of the lists after every state change, for which we have to visit all neighbors of a finite fraction of the network. The linear increase of the simulation times for the optimized algorithms  is reflecting that the amount of independent events is proportional to the number of infected vertices. Note that if the analysis is  not done close to the epidemic threshold, relatively small systems are sufficient to obtain the behavior of the thermodynamical limit and OGA  will be sufficient to this job.

\subsection{Contact process}
\label{subsec:cp_alg}

The contact process (CP)~\cite{Marrobook,Castellano:2006,Mata14}
is obtained from the generic epidemic dynamics with $\mu_i=\mu$, $\alpha_i=\infty$, 
and $\lambda_{ij}=\lambda A_{ij}/k_i$. This subtle modification
in the infection rate leads to differences with SIS model that becomes
remarkable in  networks with PL degree distributions. A central one is
that the total infection rate produced by a vertex in CP is
independent of its degree and given by $w_i=\lambda$ in contrast with
$w_i=\lambda k_i$ for SIS, see Eq.~\eqref{eq:wgen}, that leads to a
finite epidemic threshold for CP for any value of the degree exponent
$\gamma$~\cite{Castellano:2006,quenched2011,Mata14}.

The total rate of healing is the same of the SIS, given by $M=\mu
\NumberInfected$. The GA algorithm for CP follows the same steps of
SIS implementation  to build the lists $\InfectedList$ and
$\SusceptibleList$. Instead of $\mathcal{L}$, a
list $\mathcal{K}$ with the degree $\mathcal{K}_p$ of the
$p=1,\ldots,\NumberISEdges$ infected vertices connected  to each
susceptible vertex recorded at $\SusceptibleList$ and the total
infection rate transmitted along IS edges
\begin{equation}
L = \lambda \sum_{i,j=1}^{N} \frac{A_{ij}}{k_i} \sigma_i(1-\sigma_j) = 
 \sum_{p=1}^{\NumberISEdges}\frac{\lambda}{\mathcal{K}_p}
\end{equation}
are computed. Note that  $\mathcal{L}_p = \lambda/\mathcal{K}_p$. The
total rate is $R=M+L$. With probability $m=M/R$ an infected vertex is
chosen with equal chance using the auxiliary list $\InfectedList$ and
healed. With probability $l=L/R$ one susceptible vertex is chosen as
an element $p$ of $\SusceptibleList$ applying a rejection probability
$\kmin/\mathcal{K}_p$ since $\lambda_\text{max}=\lambda/\kmin$. The
time is incremented, and the lists $\NumberInfected$ and
$\InfectedList$ are updated as in the SIS algorithm.

For OGA,  the total infection rate including the phantom processes is
$W=\lambda \NumberInfected$. Since $w_i=\lambda$ and
$\lambda_{ij}=\lambda/k_i$ is independent of the target $j$, the
acceptance probabilities of both chosen vertex and target neighbor
become 1 and we obtain the widely used recipe for CP
simulation~\cite{Marrobook,Castellano:2006}: An infected vertex $i$ is
selected with equal chance using the list $\InfectedList$. With
probability $m=M/R=\mu/(\mu+\lambda)$ the selected  vertex is healed. 
With probability $w=W/R=\lambda/(\mu+\lambda)$, one of the $k_i$
neighbors of $i$ is randomly selected and, if susceptible, is
infected. Otherwise no change of state is implemented. The time
is incremented, $\NumberInfected$ and $\InfectedList$ are updated as
in the SIS algorithm.

Since the rejection method is not used in OGA, IOGA losses its sense
for CP.

\begin{figure}[tbh]
\centering
\includegraphics[width=0.95\linewidth]{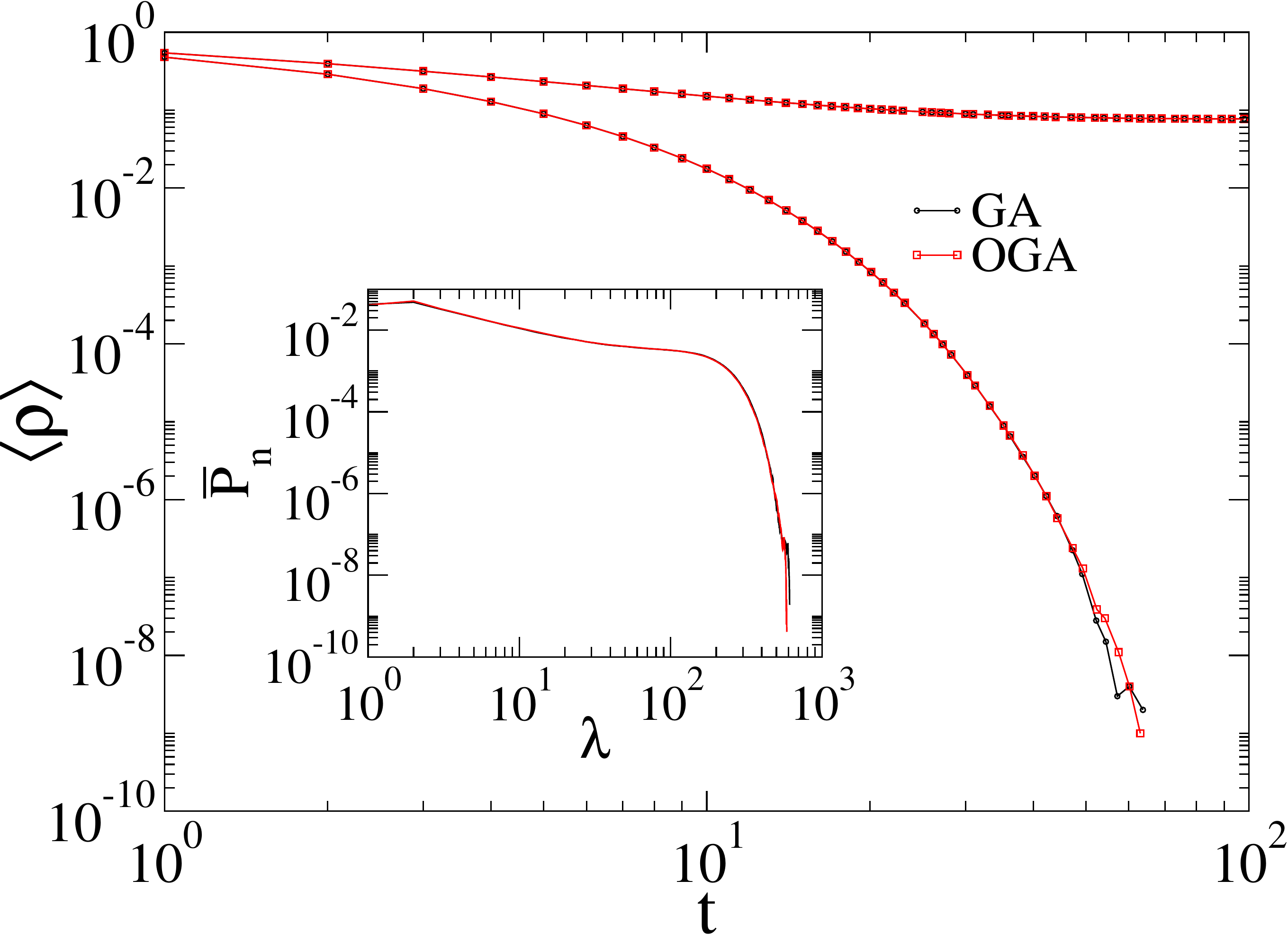}
\caption{Comparison of algorithms for CP with $\mu=1$ on an UCM network
with $\gamma=4.0$, $N=10^4$, $k_0=3$, $k_c\sim N^{1/\gamma}$ using
infection rates $\lambda=0.9<\lambda_c$ (bottom curves) and
$\lambda=1.5>\lambda_c$ (top curves). The averages were performed {over}
$10^5$ and $10^4$ samples for $\lambda<\lambda_c$ and
$\lambda>\lambda_c$, respectively. Inset shows the QS
distribution for CP dynamics at the epidemic threshold (see Table~\ref{tab:thre}) on a UCM network
with the same parameters of the main plot.}
\label{fig:decayCP}
\end{figure}

\begin{figure}[th]
\centering
\includegraphics[width=0.99\linewidth]{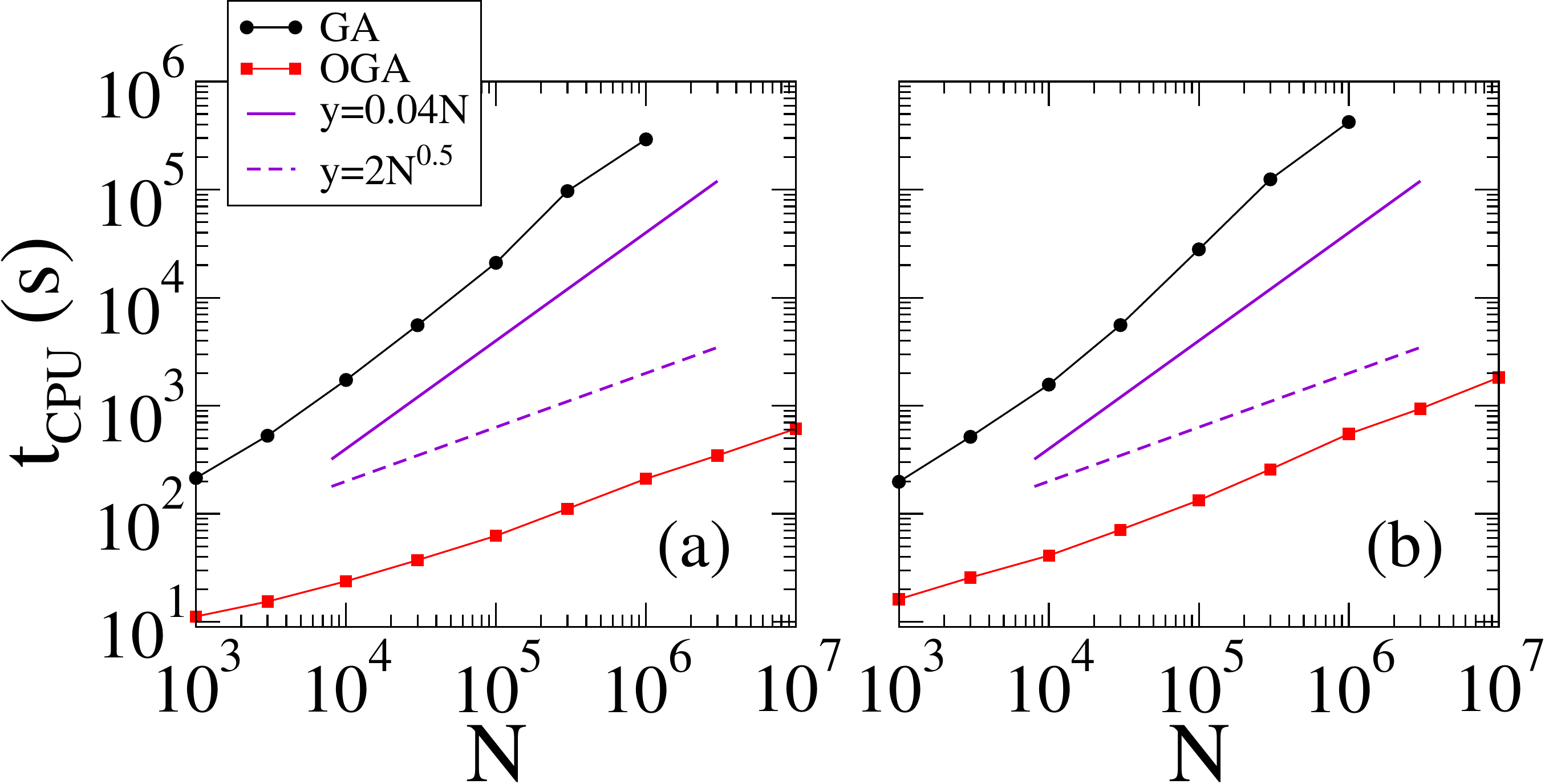}
\caption{Comparison of CPU times for QS simulations of the CP with
	$\mu=1$ using two algorithms. The simulations were performed at the
	epidemic threshold (see Table~\ref{tab:thre}) on UCM networks with minimal degree $k_0=3$,
	$k_c\sim N^{1/\gamma}$ and degree exponents (a) $\gamma=2.3$ and (b)
	$\gamma=4.0$. The total simulation time is $t_\text{av}+t_r=3\times
	10^6$. Lines are guides to the eyes.}
\label{fig:cputimeCPcrit}
\end{figure}

The equivalence between GA and OGA algorithms for CP is shown in
Fig.~\ref{fig:decayCP} for both decay and QS simulations. As in SIS,
the curves are distinguishable only at very low values due to finite
statistics. The computational times for both algorithms are compared
in Fig.~\ref{fig:cputimeCPcrit} where we see that the critical
dynamics (see Table~\ref{tab:thre} for thresholds) can be several orders of magnitude slower in the
non-optimized algorithm and the difference increases with the network
size. As in SIS, the differences become larger above the epidemic
threshold.

\subsection{SIR model}
\label{subsec:sir_alg}

Choosing  $\mu_i=\mu$, $\alpha_i=0$, and $\lambda_{ij}=\lambda A_{ij}$ for all
vertices, one obtains the SIR model~\cite{anderson92}. Differently
from SIS and CP,  SIR does not have an active stationary state. The
implementation of SIR is very similar to SIS  with the difference that
the transition I$\rightarrow$S is changed to  I$\rightarrow$R, and vertices in state 
R do not change in this model.

We performed SIR simulations starting with a single infected vertex
and the remaining of the network susceptible. To reduce fluctuations 
we always start in the most connected vertex of the network. The
simulation proceeds until an absorbing state is reached and the
averages were performed over $10^5$ repetitions in the same
network. The list of recovered {vertices} $\RecoveredList$
is not necessary for this dynamical simulation since the recovered
vertices do not have dynamics. However, it can be useful to keep this list updated and use it to setup the initial condition efficiently visiting only the recovered vertices and resetting them to the susceptible state.
A gain of up to one order of magnitude can be obtained
with this procedure since many samples do not lead to large
outbreaks, specially near and below the epidemic threshold. Thus,
after an outbreak, only a few vertices have to be updated  to reset the initial condition.

\begin{figure}[h]
\centering
\includegraphics[width=0.65\linewidth]{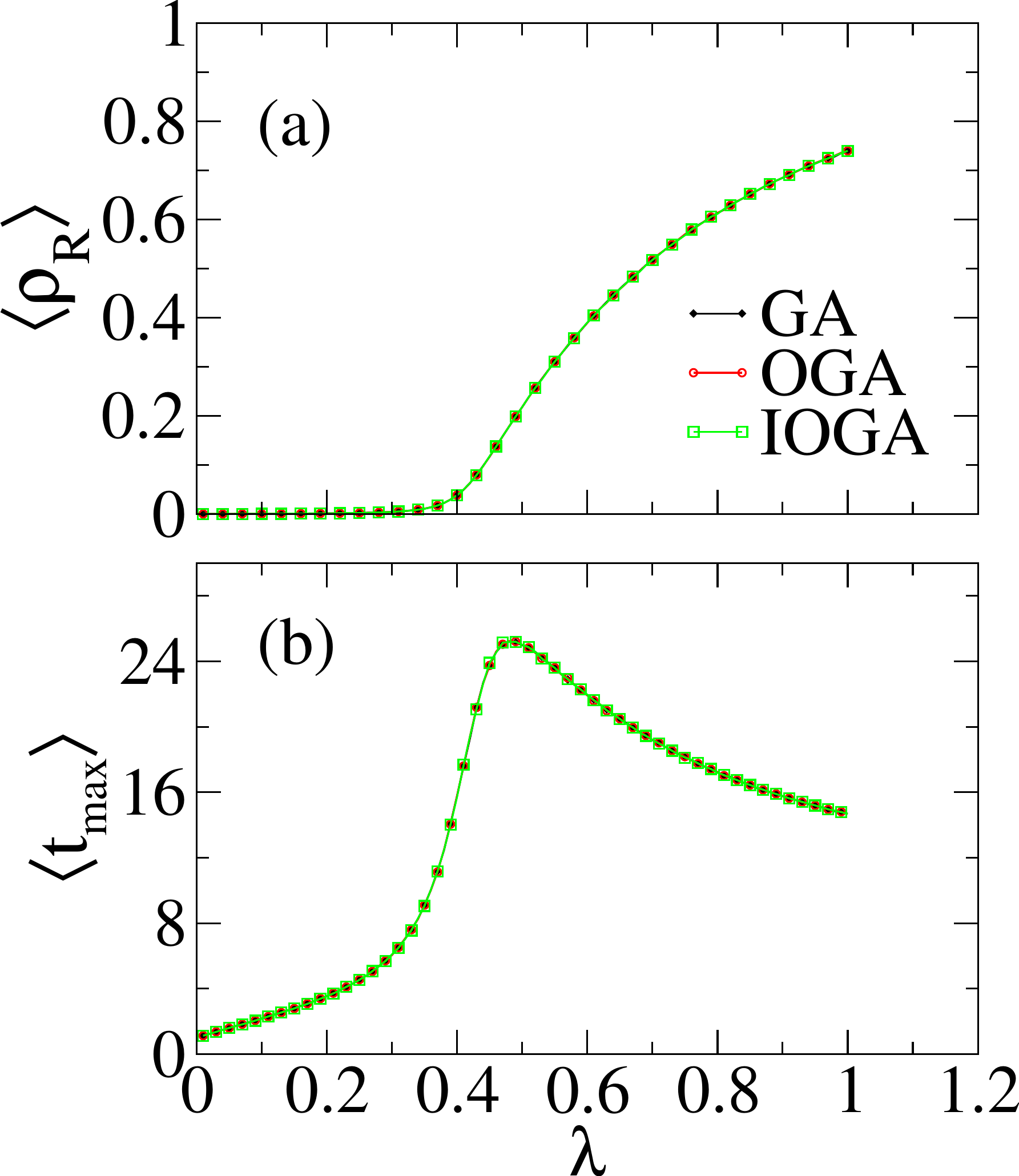}
\caption{Comparison of the SIR dynamics with different algorithms in a
UCM network with degree exponent $\gamma=4.0$ and $N=10^4$ vertices, $k_0=3$ 
and $k_c\sim N^{1/\gamma}$.
The final (a) average density of recovered vertices and (b) epidemic
lifetime are shown as a function of the infection rate. The curves
correspond to average over $10^5$ samples.}
\label{fig:comparaSIRg4p0}
\end{figure}

We calculated the final density of recovered vertices and the average
time that the activity in the epidemic outbreak lasts. The
equivalence between algorithms for SIR is shown in
Fig.~\ref{fig:comparaSIRg4p0} for a UCM network with $\gamma=4.0$.
Other degree exponents were  considered and the equivalence corroborated.
The time dependence of the average number of infected and recovered {vertices}
are also indistinguishable when the three algorithms are used. The
comparative computer efficiency of SIR algorithms is similar to SIS.

\subsection{Algorithms for more complicated dynamics}
\label{subsec:complicated}

We now provide two others examples of the {OGA} implementation. For
$\mu_i=\mu$, $\alpha_i=\alpha$, and $\lambda_{ij}=\lambda A_{ij}$ we
have the classic susceptible-infected-recovered-susceptible
(SIRS)~\cite{anderson92} model and the algorithm described in
Ref.~\cite{Ferreira16} as follows. The lists $\InfectedList$ and
$\RecoveredList$ and the variables $\NumberRecovered$,
$\NumberInfected$, and $\NumberIEdges$ (\textit{cf}. SIS algorithm for
OGA) are computed and constantly updated. The total healing, waning
of immunity and infection {attempt} rates are   $M = \mu
\NumberInfected$,  $A = \alpha \NumberRecovered$, and $W = \lambda
\NumberIEdges$, respectively, with a total rate $R=M+A+W$. With
probability $m=M/R$ one infected vertex is selected at random using
$\InfectedList$ and healed. With probability $a=A/R$, a recovered
vertex is {selected at random} using $\RecoveredList$ and converted to
susceptible. Finally, with probability $w=W/R$, an infection attempt
is performed in two steps: An infected vertex $j$ is selected with
probability proportional to its degree.  A neighbor of $j$ is selected
with equal chance and, if susceptible, it is infected. The other steps
are the same of SIS.

The generalized SIS model with $\mu_i=\mu$, $\alpha_i=\infty$ and
$\lambda_{ij}=\lambda A_{ij}/(k_i k_j)^\theta$, where $\theta$ is a model
parameter, was proposed and investigated by Karsai, Juh\'asz, and
Igl\'oi (KJI)~\cite{Karsai}. We can derive the implementation of the
KJI model presented in Ref.~\cite{Ferreira16} generalizing
the SIS algorithm using Eqs.~\eqref{eq:Qger} and \eqref{eq:wgen} as
follows. The lists $\InfectedList$ and $\mathcal{W}$ and the variables 
$\NumberInfected$ and $W$, Eq.~\eqref{eq:QgerList}, are computed and
constantly updated. The total healing rate is $M=\mu \NumberInfected$
and  the total rate is $R=M+W$. With
probability $m=M/R$ one infected vertex is selected at random using
$\InfectedList$ and becomes susceptible. With probability $w=W/R$, a vertex $i$ is chosen as an
element $p$ of $\InfectedList$ using a rejection
probability $\mathcal{W}_p/w_\text{max}$. Next, a neighbor of
$i=\InfectedList_p$, namely $j$, is selected using a rejection
probability $\left[k_i^\text{(min)}/k_j\right]^{\theta}$ where
$k_i^\text{(min)}$ is the smallest degree among the neighbors of $i$. If
$j$ is susceptible, it becomes infected. Other steps are the same
of SIS.

\section{Concluding remarks}
\label{sec:conclu}

In the present work, we show how to build statistically exact
algorithms for simulations of generic epidemic processes on very large
and heterogeneous networked systems. Grounded in the classical
Gillespie algorithm~\cite{Gillespie1,Gillespie2} for simulation of
stochastic processes, we developed optimized versions of the GA by
introducing the idea of \textit{phantom processes} that are
transitions that do not lead to changes of states but do count for
time increments. These phantom processes simplify hugely the
determination of the all possible events and the optimized Gillespie
algorithms can be several orders of magnitude more efficient that the
original one but still providing statistically exactly simulations. We
provide comparisons for the equivalence among the methods and
compared their computer efficiencies for basic epidemic models,
namely, the SIS, contact processes, and SIR models.

The original GA is much slower than OGA due to the building of the lists of all possible events after every change of state, for which is necessary to check all neighbors of each infected vertex. We could roughly estimate the number of operations per time step of GA  as of order $\NumberInfected \av{k}$ while in OGA it  is approximately constant. The number of operations by time unity is inversely proportional to the average time step and simulation CPU time  will increase with $\NumberInfected $ as well.  So, the optimized algorithms we investigated constitute great optimizations when the density of infected  vertices is low, which  is particularly relevant for analyses close to the epidemic threshold.

Searching in the literature, one can find implementations of continuous-time epidemic
models that are not statistically
exact~\cite{Klemm,Pastor01b,Odor2013b}. So, it would be interesting
to check the impact of these modified implementations on the
final outcome comparing them with the statistically exact prescriptions
described in this work. The ideas developed here can be used as a
groundwork for the building of efficient algorithms for other Markovian
dynamical process and also for building optimizations for breakthrough
topics on network theory involving non-Markovian epidemic
processes~\cite{boguna2014,Kiss15} and temporal
networks~\cite{Vestergaard2015}.

\medskip

\section*{Acknowledgements}

This work was partially supported by the Brazilian agencies CAPES,
CNPq and FAPEMIG. SCF is thankful to Romualdo
Pastor-Satorras and Claudio Castellano who were the partners in the
beginning of the learning of the Gillespie-like algorithms for SIS
model. We specially thank Romualdo Pastor-Satorras for several
suggestions that guided us towards the final shape of the manuscript
during his visits to UFV supported by program {\it Ci\^encia sem
	Fronteiras} - CAPES under Project No. 88881.030375/2013-01.

\section*{References}


\end{document}